# Phase-resolved visualization of radio-frequency standing waves in superconducting spiral resonator for metamaterial applications

A. A. Leha[1], A. P. Zhuravel[1], A. Karpov[2], A. V. Lukashenko[3], and A. V. Ustinov[2,3]

[1]*B. Verkin Institute for Low Temperature Physics and Engineering of NAS of Ukraine, Kharkiv 61103, Ukraine*
E-mail: zhuravel@ilt.kharkov.ua

[2]*National University of Science and Technology (MISiS), Moscow 119049, Russia*

[3]*Karlsruhe Institute of Technology (KIT), Karlsruhe 76131, Germany*

Superconducting microcircuits and metamaterials are promising candidates for use in new generation cryogenic electronics. Their functionality is largely justified by the macroscopic distribution of electromagnetic fields in arranged unit cells, rather than by the microscopic properties of composite materials. We present a new method for visualizing the spatial structure of penetrating microwaves with microscopic resolution in planar superconducting macroscopic resonators as the most important circuit-forming elements of modern microelectronics. This method uses a low-temperature laser scanning microscope that examines the phase (i.e., direction) and amplitude of local radio-frequency currents versus the two-dimensional coordinates of the superconducting resonant structure under test. Phase-sensitive contrast is achieved by synchronizing the intensity-modulated laser radiation with the resonant harmonics of the microwave signal passing through the sample. In this case, the laser-beam-induced loss in the illuminated area will strongly depend on the local phase difference between the RF carrier signal and the spatially temporal structure of the focused laser oscillation. This approach eliminates the hardware limitations of the existing technique of radio-frequency microscopy and brings the phase-sensitive demodulation mode to the level necessary for studying the physics of superconducting metamaterials. The advantage of the presented method over the previous method of RF laser scanning microscopy is demonstrated by the example of the formation of standing waves in a spiral superconducting Archimedean resonator up to the 38th eigenmode resonance.

## 1. Introduction

Electromagnetic metamaterial (EM MM) is an artificially engineered ($\mu_r$ negative and/or $\varepsilon_r$ negative) effective media capable of controllable the routing and manipulating EM wave propagation in a way that does not occur in natural materials. Here, $\mu_r$ and $\varepsilon_r$ are the relative permeability and relative permittivity of the medium, respectively. These capabilities can be used to create new radio-frequency (RF) devices with unique functionalities [1–8]. Experimentally confirmed RF effects include a negative refractive index [9], broadband absorbing [10], electromagnetic induced transparency [11], dense mode spectra with left-handed dispersion [12], light slowing [13], and artificial magnetism [14, 15].

A recent trend in EM MMs is the development of planar structures containing superconducting (SC) unit cells that act as so-called meta-atoms. Among them, superconducting spiral resonators (SRs) are especially attractive for the MMs research because of their magnetoactive permittivity and ultra-compact design [16, 17]. The deep sub-wavelength SC SR has been realized with a size smaller than $\lambda_0/800$ [18], where $\lambda_0$ is the free space wavelength. Additionally, superconductors have small values of the surface resistance at RF and can cause large inductance values without the associated losses found in their normal metal-based counterparts.

The RF performances of the SC SRs are often influenced by nonlinear effects at high RF power. The dominant

*A. A. Leha, A. P. Zhuravel, A. Karpov, A. V. Lukashenko, and A. V. Ustinov*

sources of these nonlinearities can be strongly localized in the dissipative region of densities of supercritical high-frequency currents generated under the standing wave crests at the edges of adjacent turns [14, 19], as well as due to the anisotropic nonlinear Meissner effect in unconventional superconductors arising both in bulk and on surface mechanisms [20]. In addition, there is a lack of understanding of the causal relationships between the spatially-temporal structure of the formed patterns of standing wave oscillations and high-frequency dynamics, including spatial attenuation, blocking, regrouping, and redirection of propagating microwaves. Therefore, a microscopic technique for spatially resolved identification of the phase relationships of resonant oscillations of individual meta-atoms is important for clarifying these unexplored features. For this purpose, the contrast of the image of the distribution of the RF field was previously studied in detail using a laser scanning microscope (LSM) at various electromagnetic modes of the spiral [21–23].

Currently, available RF LSM methods provide information on the local amplitude of microwave resonances, completely ignoring the contribution of the vector directions of RF currents and visualizing only their amplitudes [23–27]. To probe the spatial distribution of the laser-beam-induced RF photoresponse in usual LSM modes of the RF imaging contrast, a modulated laser beam is focused onto and scanned over the surface of an SC SR. The image contrast of these changes is recognized in terms of the square of the local RF current densities, which is *a priori* scalar quality. Clearly, this measurement alone does not allow visualization of the RF current flow direction over probed geometry due to inadequate data on the phase of the electric field distribution.

To avoid the problem of incorrect phase presentation in the RF LSM response, we apply a recently developed RF LSM regime [28]. It used synchronous demodulation of the spatial form of a standing wave, perturbed by a weak action of a scanning laser probe, the oscillations of which were synchronized with the microwave carrier signal of a spiral superconducting resonator. The bottleneck of this LSM detection circuit was the use of a crystal diode, which, when nonlinearly mixing microwave signals, generates a huge spurious component of the CW amplitude.

In this study, we deviate from the traditional RF LSM imaging method proposing a more direct method for phase-sensitive measurements, thereby demonstrating the ability of superconducting resonators to behave as natural nonlinear microwave mixers. The general idea is to synchronize signals of the RF drive and a laser beam modulation, adjusting them in such a way as to tailor the phase-sensitive photoresponse. In order to present imaging LSM contrast of the vector RF current density in terms of both phase and amplitude of oscillating components, the nonlinear mixed signals are a heterodyne detected in the function of the laser probe position ($x, y$). This approach removes the frequency limitations of the existing RF LSM regime and brings the phase-sensitive LSM regime to the level required for exploring the physics of superconductive metamaterials.

We demonstrate the feasibility of LSM imaging the magnitude and phase components of a RF current density vector field. As an example, radial magnitude and phase profiles of the concentric circular patterns of RF standing waves in the superconducting spiral resonator are imaged for the first three resonant modes. We performed the time-domain and frequency-domain simulations of the RF signal propagation and found an excellent agreement with the obtained experimental data. The developed experimental technique allows to image the internal electrodynamics of SC MMs and provides new opportunities for their characterization.

We believe that the creation of an LSM image with phase sensitivity in this new way will make a significant contribution to the development of experimental devices for studying the spatial structures of penetrating microwaves — one of the most important problems in understanding the mechanisms of microwave response of superconducting metamaterials. These efforts are aimed at finding a new experimental approach to a detailed understanding of the unusual physics, technological limitations, and characteristics of radio-frequency devices based on MMs.

## 2. Samples and measurement technique

### 2.1. Spiral resonator

In this article, we consider an RF magnetic metamaterial, the design of which is based on a planar Archimedean spiral resonator (ASR). Since the design is realized with superconducting film, the structure has a low surface resistance $R$ and significant values of inductance $L = L_{geo} + L_{kin}$ of a multi-turn SC coil. The presence of an intrinsic capacitance $C$ in the turn-to-turn gaps makes such a spiral self-resonating meta-atom for use in the frequency range below 100 MHz. Here, $L_{geo}$ is the geometric inductance and $L_{kin}$ is the kinetic inductance of an equivalent $RLC$ circuit describing the frequency of the first fundamental half-wavelength eigenmode:

$$f_1 = \frac{1}{2\pi}\sqrt{\frac{1}{(L_{geo}+L_{kin})C}} \quad (1)$$

and its unloaded quality factor as $Q_u = (1/R)\sqrt{LC}$. The lumped $RLC$ parameters of the ASR are linearly scaled with the arch length $l_{MS}$ of a microstrip inductor as

$$l_{MS} = \frac{S}{4\pi}\left(\gamma\sqrt{1+\gamma^2} + \sinh^{-1}\gamma\right)\Big|_{\gamma_0}^{\gamma_i}, \quad (2)$$

where $\gamma_0$ and $\gamma_i$ are the polar angles for the outer $\rho_0$ and inner $\rho_i$ radii of a spiral-filled ring, respectively.

In this case, the distribution of the standing wave currents $J_{RF}(l_g)$ along the longitudinal direction $l_\gamma$ on a resonating strip is well approximated as those of a one-dimensional microstrip waveguide with open boundary conditions at two ($\rho_0$ and $\rho_i$) identical open ends [29–31]:

$$J_{RF}(l_\gamma) = J_0 \sin(n\pi l_\gamma / l_{MS}), \qquad (3)$$

where the peak current density at $l_g = l_{MS}/2$ is given by [32]

$$J_0 = \frac{1}{Wt}\sqrt{\frac{r(1-r)8Q_u P_{RF}}{n\pi Z_0}}, \qquad (4)$$

here $r$ is the voltage insertion loss obtained from the transmission characteristic, $n$ is the index of the harmonic mode, $P_{RF}$ is the circulating RF power, and $W$, $t$, and $Z_0$ are the width, the thickness, and the characteristic impedance of the microstrip transmission line, respectively.

A simplified model of the $J_{RF}(l_\gamma)$ distribution [see Eq. (3)], makes it possible to clearly represent it in the form of amplitude and phase profiles of a standing wave of a one-dimensional microstrip wrapped in a spiral. This, in turn, opens up the possibility of easy modeling and calculation of radial (along *R*-scan direction, as shown in Fig. 1) $J_{RF}(R)$ distributions, which greatly facilitates the interpretation of the results of the study of standing waves using the RF LSM.

Figure 1 (a) shows a top view of the ASR ring geometry, commonly used in the proposed studies. Darker areas in this microscopic contrast indicate Nb metallization, while brighter areas are associated with a low surface reflectance of a 300 μm thick quartz substrate. The detailed structure of the ASR metallization is depicted in Fig. 1(b). It is made by continuously winding a $t = 200$ nm thick Nb microstrip from the inner to outer radii and has the following design: $N = 44.5$ turns, outer radius $\rho_0 = 3.0$ mm, inner radius $\rho_i = 2.1$ mm, the width of the Nb microstrip $W = 10$ μm, step size between the adjusted turns $S = 20$ μm, and the total length of the Nb multiturn microstrip as long as $l_{MS} \sim 657$ mm. The critical temperature of superconducting Nb, determined by resistive dc and transmission RF measurements, is $T_c \sim 9.2$ K.

More details about the other studied samples and their specific patterning processes can be found elsewhere [18, 34] or appears as needed in the text.

### 2.2. Global transmission data

The forward (from input port 1 to output port 2) measurements of RF transmission $S_{21}(f)$ are performed with an Anritsu MS4640A Vector Network Analyzer (VNA) at temperatures from 4.5 to 300 K. A singe ASR sample was mounted on a cooled sapphire disk in a vacuumized cavity of an optical cryostat between two nonresonant magnetic loop probes coaxial with each other and with the sample. The location of these elements and their mutual inductive coupling are sketched in the inset in Fig. 2. The inner diameter of the exciting circuit was chosen to be 1 mm larger than the outer diameter of the sample providing whole optical access to the laser probe for RF LSM scanning. This hardware limitation also forced to bring this loop closer to the sample at a distance of 3 mm, which led to their strong mutually inductive coupling. Despite the creation of an inhomogeneous magnetic field [21] and excessive coupling in the first few modes [34], this effect, nevertheless, expanded the measurement capabilities up to 10 GHz, where the efficiency of the probe circuit decreases due to dissipative losses.

The main panel in Fig. 2 shows the *global* variations of the transmission scattering coefficient $S_{21}(f)$ of the ASR sample in the frequency domain from 100 kHz to 1 GHz. This data describes the response of the linear Meissner phase at $T = 4.5$ K, well below $T_c$ and $P_{RF} = -10$ dBm much smaller than critical RF power $P_c = 14$ dBm. As evident from the graph, the frequency $f_1$ of the first ($n = 1$) fundamental mode is only 75 MHz, followed by higher modes $f_n \approx (2n - 1)f_1$, where $n = 1, 2, 3 \ldots, N$ is the index of the resonant mode. The resonant peaks of the highest modes decay rapidly with increasing excitation frequency, while even peaks are usually weaker than odd ones. All these effects are completely predictable within the framework of analytical computations proposed recently in Refs. 19, 21. Once resonant frequencies were confidently detected, they were used for RF LSM experiments.

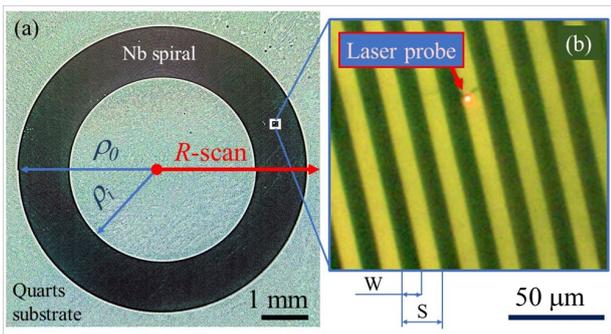

*Fig. 1.* (Color online) An optical microscope image of (a) a ring-shaped superconducting Archimedean spiral resonator (the metallized coating looks darker than the substrate) with $\rho_0 = 3$ mm, $W = 10$ μm, $S = 20$ μm, $N = 44.5$ turns and (b) the detailed view of a small outlined area that is acquired by a CCD camera incorporated in the LSM optical train. The position of the laser probe is labeled. The probe is moved along R-scan direction while probing the section profiles of RF standing wave patterns.

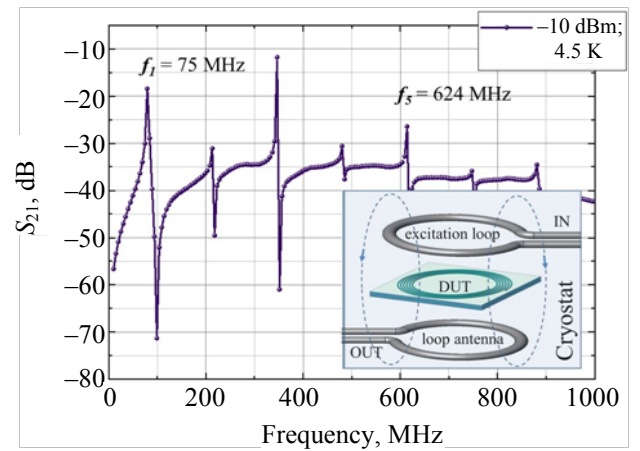

*Fig. 2.* (Color online) Transmission ratio $|S_{21}|$ vs. frequency on the Nb/quartz spiral sample showing the fundamental ($f_1 = 75$ MHz) up to 7th highest harmonic resonances at $P_{RF} = -10$ dBm and $T = 4.5$ K. Notations: RF IN is the input excitation loop, RF OUT is the output loop antenna, DUT is the ASR device under test.

### 2.3. Laser scanning microscope and its photoresponse at microwaves

Figure 3 illustrates a generalized scheme for two LSM variants capable of realizing phase-sensitive and scalar detection modes of microwaves distribution in superconducting resonant circuits. In this diagram, a galvanic jumper in the 0-2 position connects the amplified receive antenna signal to a (Schottky barrier) crystal diode that acts as an envelope detector in previous versions of the LSM setup [25–28]. By switching the microwave contactor to the 0-1 position, the RF antenna currents induced by the sample magnetic fields are directly used to synchronously detect the microwave field change in an upgraded LSM configuration. However, in either of these acquisition modes, the general LSM operation remains unchanged. It is based on the procedure of spatially-temporal sampling and reconstruction of the high-frequency field by step-by-step raster scanning of its spatially inhomogeneous structure using an active source of local surface action — the so-called laser probe. In this case, the effect of local interaction between the RF field and the probe is called a photoresponse (PR) signal. This is detected during scanning as a function of the probe position $(x, y)$. The resulting 2D PR $(x, y)$ map is converted to a 2D digital data set and stored in computer memory to produce images with the desired RF contrast.

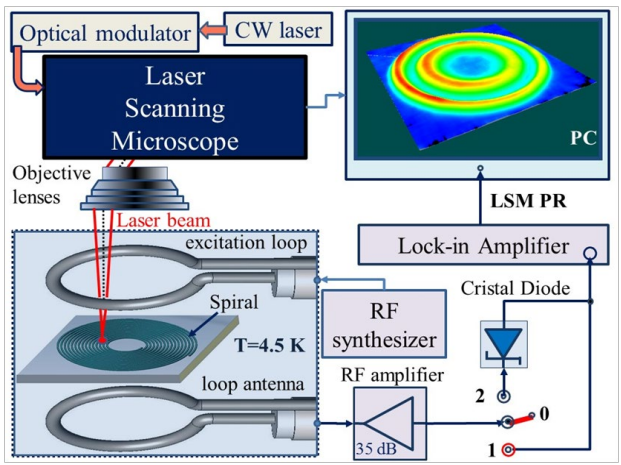

*Fig. 3.* (Color online) Simplified schematic of the RF LSM setup. Numbers 1 and 2 show the contacts of the RF switch for changing the detection mode.

In our experiments, the laser beam of a CW diode laser (Coherent, OBIS 660 nm LX 100 mW) is focused by LSM optics onto the surface of the spiral under study. An LSM objective lens (Mitutoyo Plan Apo) with 5× magnification produces a probe size of about 5 μm in a field of view up to 1 mm, while the $f$-theta lens with 20 μm focusing capability is used to obtain for large-scale LSM images covering areas of up to $10 \times 10$ mm$^2$. The intensity of the probe is an amplitude (AM) modulated in the frequency range from CW to 5 GHz using a fiber-optic Mach–Zehnder amplitude modulator (Jenoptik, AM-635HF). This, in turn, leads to oscillating local heating $\delta T(x, y)$ and direct Cooper pairs breaking under the influence of a laser light quanta, whose energy is about 2 eV, significantly exceeding the doubled energy of the superconducting gap of niobium close to $2\Delta_{Nb}(4.5\ \text{K}) \approx 5$ meV.

All LSM experiments were carried out at a fixed temperature $T = 4.5$ K, much lower than $T_c$ of the SC Nb microstrip. In this case, the built-in cryogenic arrangement of dual-loop RF excitation/antenna circuit has been practically the same as in the global RF experiment, but the sample was stimulated by an RF synthesizer (Anritsu MG37022A), as shown in Fig. 3.

While the RF currents oscillate at a fixed frequency $\omega_n = 2\pi f_n$ in the $n$th standing wave mode, the sample is disturbed by a modulated laser probe, the power of which varies as

$$\delta P_L(x,y) \approx A_F(x,y)(1-\Re(x,y))P_L^0(1+e^{i\omega_M t}), \quad (5)$$

where $P_L^0$ is the power of CW laser beam, $\Re$ and $A_F$ are the local optical reflectance and absorbance of the sample surface underneath the probe. The resulting localized exposure causes optically induced changes in the local electrodynamic properties of the SC material. These changes lead to tuning the resonant frequency and/or degradation of the $Q$ factor of the DUT. Both effects, in turn, alter the overall transmission response $S_{21}(f)$ of the device while the RF LSM technique probes the $x$–$y$ distribution of the LSM PR signal. This is detected by an inductively coupled output loop antenna that converts the photoresponse to an RF voltage signal such as

$$PR(x,y) \propto A_{SW}(x,y)S_{RF}(t)\frac{\partial |S_{21}(f)|}{\partial P_L}\delta P_L(x,y), \quad (6)$$

where $A_{SW}$ is the standing waveform, $S_{RF} = A_{RF}\sin(\omega_{RF}t + \varphi_{RF})$ is the waveform of the carrier wave, and $A_{RF}$, $\omega_{RF} = 2\pi f_{RF}$, and $\varphi_{RF}$ are its amplitude, frequency, and initial phase, respectively.

It clearly follows from the phenomenological meaning of Eqs. (5) and (6) that the SC resonator exhibits the property of a natural AM mixer for the carrier frequency of microwave pumping and modulation of the laser action.

In this case, the equation describing the stimuli can be generalized to a simplified form of nonlinear AM mixing

$$PR(x,y) \propto A_{SW}(x,y)$$
$$\times [1 + m \sin(\omega_M t + \varphi_M)]\sin(\omega_{RF} t + \varphi_{RF}), \quad (7)$$

where $m = A_F(x,y)(1-\Re(x.y))P_L^0 / A_{RF}$ is the modulation index and $\varphi$ denotes the phase difference. As follows from Eq. (7), the local standing wave magnitude $A_{SW}$ and the modulation index $m$ are changed during the LSM scan. In this case, the spatial variations of the modulation index are associated only with the enhancement of the PR with increasing laser power and demonstrate the disappearance of modulation signals only when the substrate is probed, where $A_F(x,y)$ tends to zero. Again, at each point of the LSM scan, the weighting contribution of the standing wave $A_{SW}(x,y)$ does not affect the nature of the RF oscillations in the probing region, where it can always be modeled, as shown in Fig. 4. This approach turns out to be convenient for modeling and interpreting LSM PR using standard microwave signal tracing procedures, including the fast Fourier transform (FFT) spectral domain, as shown in Fig. 5.

The demodulation process on the LSM PR signal recovery always requires a nonlinear operation on this signal to extract its message proportional to the modulation of the carrier. For the presented in Fig. 4 event of $\omega_{RF} \gg \omega_M$, the method of rectification (nonsynchronous detection) is most often used. It takes advantage of the Schottky diode circuit [28] as an envelope detector (red line in Fig. 4), the response of which is then amplified synchronously by a lock-in technique, as shown in Fig. 2, with the 0-2 switch shorted. In this regime, the diode acts as a power detector (which is, of course, phase insensitive) since the integration is not done exactly for one (or an integer) oscillation.

In this publication, we propose an alternative, not previously used, approach for direct reconstruction of the PR

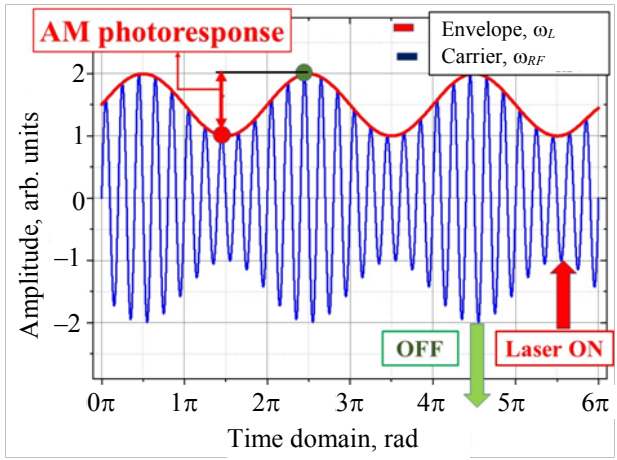

*Fig. 4.* (Color online) Time-domain example of an amplitude-modulated signal with a carrier frequency $\omega_{RF} = 75$ MHz, a modulating frequency $\omega_M = \omega_L$ 7.5 MHz, and a modulation index $M = 0.5$.

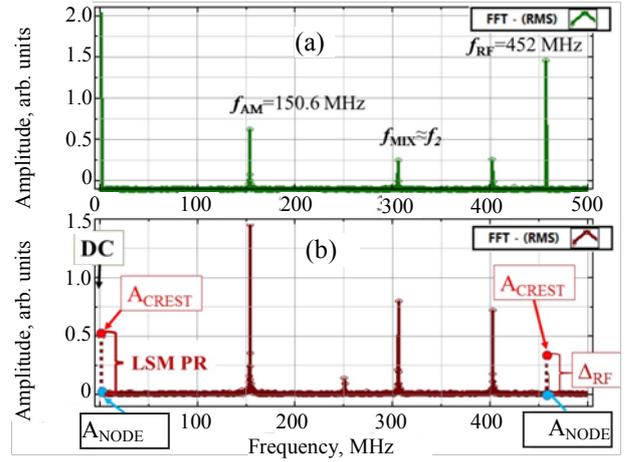

*Fig. 5.* (Color online) (a) Simulated FFT spectrum of the LSM PR input signal at the lock-in amplifier input and (b) the post-mix signal arising from a 452 MHz carrier modulated with 150.5 MHz tone.

signals that eliminates the intermediate diode circuit while simplifying the RF LSM structure by shorting the switch to the 1-0 position (see Fig. 3). An important purpose of such an LSM data acquisition system is phase-sensitive demodulation, which is capable of recovering information about the amplitude and phase of the received signal. Then we test both of these detecting systems and compare their results to be convincing.

### 3. Results and discussion

As a starting point for phase-sensitive RF–LSM measurements, it was necessary to create $PR(x,y)$ reference images for comparison. Following the procedure originally developed in Refs. 33, 35, we used proven methods of spatially resolved phase-insensitive reconstruction of the insertion loss component, $PR_{IL}(x,y)$, which can be represented in the following form:

$$PR_{IL}(x,y) \propto P_{RF} \frac{\partial[S_{21}(f=f_n)]^2}{\partial(P_L)} \delta P_L$$
$$\propto \frac{2S_{21}(f_n)}{1+4Q^2(f/f_n-1)^2} \delta P_L. \quad (8)$$

In this case, the LSM mode switch was set to the 2-0 position (see Fig. 3) to connect the crystal diode to a circuit that demodulates the low-frequency $f_M$ oscillations at a precisely fixed carrier frequency $f_n$, tuned to the transmittance peak $S_{21}(f_n)$ of the $n$th resonant mode. Figure 6 shows LSM $PR_{IL}$ in 1st, 3rd, 5th, and 38th harmonic standing wave patterns over a 7×7 mm² area. Scanning was performed by a laser probe with a diameter of $d_L = 25$ μm at a temperature $T = 4.5$ K and an RF pump power $P_{RF} = -10$ dBm on the excitation loop. As expected from theoretical predictions [19, 28, 36], the voltage response of the crystal diode restores the resonant structure with a quality $PR_{IL}(x,y) \propto \delta(\int R_s \lambda^2 J_{RF}^2(x,y)dS$, where the integral is carried out over the area $S$ of laser-induced perturbation. This LSM PR component is a convolution of the surface resistance changes $\delta R_s$, weighted by the square of the local value of the RF current density, $J_{RF}^2(x,y)$.

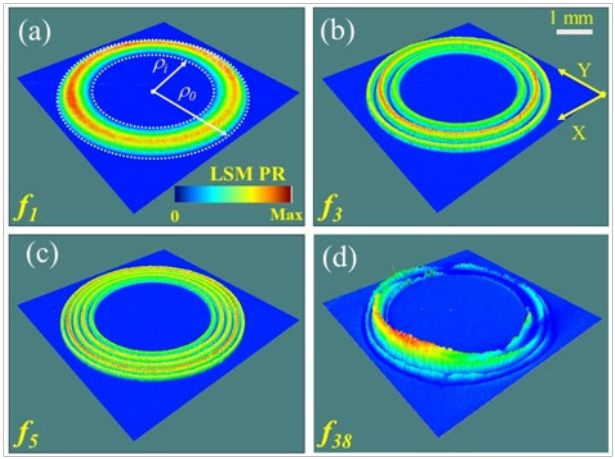

*Fig. 6.* (Color online) Pseudo-color laser scanning microscopy images showing the distribution of the RF standing wave current in a superconducting niobium spiral with $N = 40.5$ turns corresponding to resonant transmission modes at (a) $f_1 = 75$ MHz, (b) $f_3 = 354.6$ MHz, (c) $f_5 = 624$ MHz, and (d) $f_{38} = 5.6$ GHz. The dashed line in (a) outlines the edges of the spiral within radii $\rho_0$ and $\rho_i$.

In this interpretation, at least for the first few resonant *n*th modes, the spatial distribution $PR(x, y)$ depicts an axially symmetric structure of standing half-waves, the number *n* of which increases in direct proportion to the index of the excitation mode, as can be seen from Figs. 6(a)–(c). For example, any radial profile of spiral excitation at the fundamental frequency of ~ 75 MHz [Fig. 6(a)] is well described by one half-wave of a sinusoid between the radii and, taking into account, the difference in the length of the turns during winding. Although the standing waves of higher [such as in Fig. 6(d)] resonant modes are becoming less and less predictable for perception within the framework of the proposed geometric approach, we still managed to show that they, too, with some success fall within the description of this simple model.

A two-dimensional version of Fig. 6(d) is shown in Fig. 7(c) in grayscale contrast for visual convenience. The brightest areas correspond to the highest RF current densities, while their zero values appear almost black. Here, the red/blue orthogonal lines show the *x*/*y* diametrical scan directions of the reference lines to represent the LSM PR in the form of profiles shown in Figs. 7(a) and 7(b). The contrast of these images clearly shows the LSM PR anisotropy of the 38th resonant mode response, which was not observed at the lowest harmonics. For a spatial detailing of the response, the profile from the area colored red in Fig. 7(b) extends through the turn section as shown in Fig. 7(e). An example of numerical simulation of the response in this section is shown with turn-by-turn sampling in the conjugate line-scan coordinate of Fig. 7(e). There is a good qualitative agreement between both data types belongs to the central turns of the spiral.

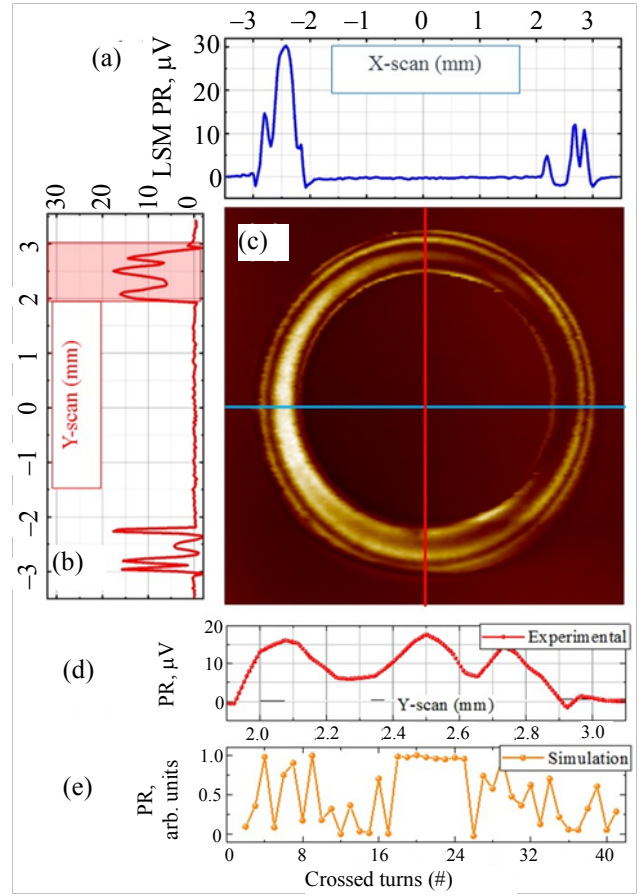

*Fig. 7.* (Color online) The profiles of the PR amplitude obtained in orthogonal cross-sections of diametrical scanning along the *x* (a) and *y* (b) coordinates of the standing wave for the 38th resonant mode of ACR are shown in (c) by two-dimensional brightness image. The semitransparent red rectangle in (b) outlines the RF LSM PR of those turns whose PR is detailed in (d) together with its (e) numerical simulations.

Based on the data presented in Fig. 7, several generalizing conclusions can be drawn at once. First, the higher harmonics of the resonance cause the loss of axial symmetry of the RF field distribution in the multiturn spiral, as can be seen from Fig. 7(c). In this sense, there is a tendency towards surface polarization of localized electromagnetic waves. when the spiral no longer obeys the advantages of a lumped atom in the effective medium of a metamaterial. Second, for an *N*-turn ring resonator operating in the *n*th resonant mode, the maximum anisotropy is observed under a condition close to $n \sim N$ if we assume that the mode structure is not caused by other unknown reasons for the spatial self-organization of the wave patterns. It is intuitively predictable that the effect is associated with the spatial grouping of the *n*th number of half-waves at the excitation frequency $f_n$, when each standing half-wave has the size of the length of each $n = N$ spiral turn. Third, comparing the profiles in Figs. 7(a) and 7(b), we can see that the used simple model of winding standing waves in a spiral [see Eqs. (2) and (3)] adequately describes the behavior of the LSM PR obtained from experimental observations. The discrepancy between these data, rather, is due to the difference in the procedure of their presentation. The experimental LSM profiles [Fig. 7(d)] were measured with a scan step of about 5 μm using a 25-μm probe that effectively averages the response spatially by probing several turns simultaneously. Whereas the simulated LSM profile [Fig. 7(e)] displays the distribution of the normalized current density squared over the turns with a scan resolution that exactly matches the 20-μm step size between adjusted turns. At the same time, in modeling, important details about phase changes in the spatial structure of standing waves are masked when the PR amplitude is squared. All this can lead to an inaccurate assessment of the agreement between experimental and calculated data, the problem of which we tried to clarify in subsequent experiments.

A standing wave of the second resonant mode can serve as the simplest, but at the same time, a well-recognized object for elucidating the ability of LSM to visualize spatially localized structures of in-phase/anti-phase microwave oscillations over the area of a superconducting resonator. As it was proved by the experience of previous studies [14, 21, 37], the stationary structure of such a pattern in a ring-shaped spiral appears as two concentric standing sinusoidal half-waves between its inner and outer diameters, separated by a single circular nodal line with zero oscillations. Figure 8(a) shows an illustrative example of such a pattern obtained in the diode detection mode of the response (the mode switch is in the 0-2 position in Fig. 3). This is a quasi-three-dimensional large-scale (7×7 mm$^2$) image of a standing wave of the second resonant mode ($f_2 = 220$ MHz) in Nb ASR, obtained with a laser modulation frequency of 1 MHz at a $T = 4.5$ K and a power of 10 dBm. In the image, a line with an arrow indicates the direction of the radial scan along which the phase-sensitive LSM photoresponse is measured when the switch in the RF measurement circuit in Fig. 2 is reset to 0-1 position. As expected, when the recording channel was switched to the phase-sensitive measurement mode, in the spatial structure of two standing half-waves associated with the countercurrent of screening RF currents on opposite sides of the nodal line of the standing wave, LSM PR of opposite sign appeared from reversed-phase action.

Figure 8(b) compares the measured and simulated $PR_{IL}(\rho)$ profiles plotted along the line-scan and under the same condition that has been used in Fig. 8(a). Experimental data (blue curve) were LSM acquired at fixed temperature $T = 4.5$ K (well below $T_c = 9$ K), $P_{RF} = 14$ dBm (below $P_c = 18$ dBm) and $f_M = 1$ MHz (well below $f_2 = 220$ MHz). These show the $PR_{IL}(\rho)$ of individual turns probed between radial positions $\rho_i$ and $\rho_0$ with a 5-μm spatial resolution. The modeled (red line) curve is the envelope of the standing wave, reconstructed using Eq. (6) in terms of the $J_{RF}^2(\rho)$ radial distribution. There is good agreement between the data, except for the position of the ASR edges and nodal turns.

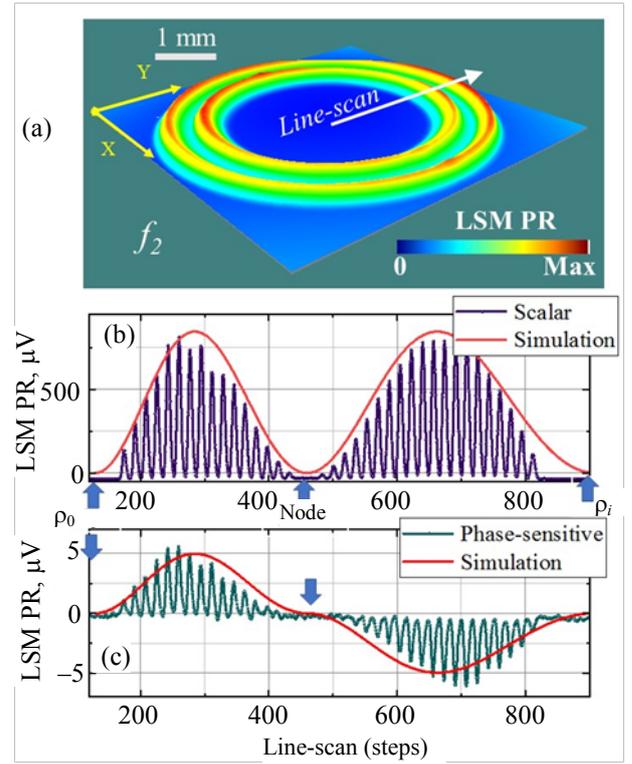

*Fig. 8.* (Color online) (a) False color 3D image of LSM PR(*x*, *y*) showing RF current-induced dissipation in a Nb spiral at 4.5 K, 14 dBm, and at the second harmonic ($f_2 = 220$ MHz). The line-scan direction indicates just radial position was used to plot the PR-section profiles investigated in (b) phase-insensitive "scalar" and (c) phase-sensitive imaging contrast regimes.

The same effect is observed with phase-sensitive measurements presented in Fig. 8(c). This manifestation is most likely associated with the predominant contribution of the nonequilibrium component to the LSM $PR_{IL}(\rho)$ described by Eq. (8). It predicts that the first detectable insertion losses in the resonant transmission $[S_{21}(f = f_n)]$ can be presented as

$$PR_{IL}(\rho) \propto [S_{21}(P_{RF}, f_n)]^2 - [S_{21}(P_{RF} + \delta P_L, f_n)]^2 \quad (9)$$

for a condition $J_{RF} \geq J_c(\rho) - \delta J_c(\rho, P_L)$ combining the impacts of the local microwave excitation (first term) and suppression current by the laser energy (second term). It has been described earlier in [23, 36]. With regard to local photoinduced changes in surface resistance $\delta R_s(x_0, y_0)$, $PR_{IL}$ is directly related to Ohmic dissipation generated by the laser probe at position $(x_0, y_0)$:

$$PR_{IL}(x_0, y_0) \propto J_{RF}(x_0, y_0) \delta R_s(x_0, y_0). \quad (10)$$

Within the framework of the phenomenology described in [38] (in the case of a linear response function and a weak perturbation by the probe) for the geometry of a microstrip oriented along the path $l_{MS}$, the change in $\delta R_s(x_0, y_0)$ as a function of the local critical current $J_{RF} \geq J_c(l_{MS}^0, P_{RF}) - \delta J_c(l_{MS}^0, P_L)$ at a particular position of the laser probe $l_{MS}^0(x_0, y_0)$ can be described as

$$\delta R_s(l_{MS}^0) \propto \frac{\pi}{4} \frac{\Lambda}{Wl_{MS}} \int_{L_{MS}} dL_{MS} \frac{\partial R_s}{\partial J_c(l_{MS}^0)}$$

$$\times \frac{\partial J_c(l_{MS}^0)}{\partial P_L}|_{P=P_{RF}+\delta P_L} \delta P_L(l_{MS}^0) \quad (11)$$

for a large-scale imaging ($\Lambda \geq W$), where $L_{MS}$ is the path along the entire spiral with a total length $l_{MS}$, and $\Lambda$ is the characteristic decay length, describing the spatial attenuation of $PR_{IL}(l_{MS}^0) \propto e^{-|l-l_{MS}^0|/\Lambda}$ at a distance $l$ out of the beam focus. One can assume that both terms $\partial R_s/\partial J_c$ and $\partial J_c/\partial P_L$ do not change under the probe if the probe diameter $d_L$, $\delta P_L$, and $\Lambda$ are spatially homogeneous over the entire resonator area. By combining Eq. (10) with the integrand of Eq. (11), it can be seen that inside the laser probe

$$PR_{IL}(l_{MS}^0) \propto \frac{\Lambda}{Wl_{MS}} J_{RF}^2(l_{MS}^0) \frac{\partial R_s(l_{MS}^0)}{\partial J_c(l_{MS}^0)}|_{J=J_{RF}}$$

$$\times \frac{\partial J_c(l_{MS}^0)}{\partial P_L}|_{P=P_{RF}+\delta P_L} d_L \Lambda \delta P_L(l_{MS}^0), \quad (12)$$

along with the polar coordinate $L_{MS}$ in the quasi-one-dimensional geometry of the microstrip. Note that Eq. (12) shows the threshold $R_{IL}$ generation mechanism for both $P_{RF}$ and $\delta P_L$ excitations. In the undercritical state of the superconducting structure at $P_{RF} + \delta P_L \leq P_c$, the value of $\partial R_s$ is zero for any position of the laser probe, where $P_c$ is the critical RF power. In this case, the $PR_{IL}$ cannot be detected by LSM at any $f_n$ in the insertion loss contrast imaging mode, which is a limiting factor in its use for elucidating the real spatial structure of penetrating microwaves.

In addition, when switching modes from a crystal diode circuit (see position of jumper 0-2 in Fig. 3) to direct (see position of jumper 0-1 in Fig. 3), antenna detection more than a hundredfold degradation in the LSM PR is observed, as can be seen from comparisons of Figs. 8(b) and 8(c). This, in turn, significantly reduces the signal-to-noise in the lock-in amplifier circuit and leads to the need to increase the LSM image capture time. To eliminate this problem, it would seem more optimal to develop a distinctive method for the generation of phase-sensitive LSM signal by exploring the heterodyne mechanism of mixing the laser modulation frequency with microwave frequency. For this, it is desirable to use the phase-sensitive detection mode for the inductive LSM PR component well below $T_c$, using the ideas for improving the useful response signal described in [26–28]. The implementation of this idea is currently under development. We hope that these efforts will remove the existing restrictions on frequency and noise and bring the phase-sensitive LSM to the level necessary for studying the physics of superconducting metamaterials.

## Conclusions

This article presents the first preliminary experimental results on using the RF LSM technique to the 2D phase-sensitive imaging spatial distribution of microwave currents screening penetration and guiding the propagation of microwaves in a superconducting magnetic metamaterial. It was shown that a unique image contrast of localized insertion loss is created directly in the resonant circuit of a superconducting meta-atom as a result of nonlinear AM mixing of the propagated RF carrier signal with in-phase synchronized local excitation of a focused laser beam.

The presented RF LSM method is used to study standing waves in a superconducting spiral Archimedes resonator up to the 38th eigenmode of its resonance. An enhancement in the anisotropy of Meissner screening currents with an increase in the eigenmode index of the resonant excitation is found. This effect can be useful for creating subwavelength structures similar to atomic lattices, such as waveguides and polarizers, capable of manipulating, redirecting, and controlling the propagation of radio-frequency waves.

To confirm the effectiveness of the phase-sensitive method, we demonstrated and described the results obtained for the second standing wave pattern in the scalar LSM PR demodulation regime and compared them with phase-sensitive measurements. The obtained data are in excellent agreement with the simulation results. The proposed technique for phase-sensitive imaging of the microwave currents with micrometer-scale resolution opens the way to explore and understand a great variety of complex RF current distributions and microwave patterns in superconductive circuits. Despite the fact that the developed method made it possible to study the amplitude and direction of microwave signals in a meta-atom, it is not ideal and requires improvement work. Our further work will be aimed at creating a heterodyne mechanism for mixing a laser beam and a microwave signal in a sample to increase the signal-to-noise ratio.

## Acknowledgments

We acknowledge valuable contributions from Steven M. Anlage and Cihan Kurter for making high-quality samples. This work was partially supported by the Volkswagen Foundation (Grant Az97768).

___

Фазово-чутлива візуалізація радіочастотних стоячих хвиль у надпровідному спіральному резонаторі для застосування в метаматеріалах

A. A. Leha, A. P. Zhuravel, A. Karpov, A. V. Lukashenko, A. V. Ustinov

Надпровідні мікросхеми та метаматеріали — перспективні кандидати для використання в кріогенній електроніці нового покоління. Їх функціональність значною мірою визначена макроскопічним розподілом електромагнітних полів в системі структурних елементів, ніж мікроскопічними властивостями композитних матеріалів. Представлено новий метод візуалізації просторової структури проникальних мікрохвиль з мікроскопічною роздільною здатністю в площинних надпровідних макроскопічних резонаторах, що є найважливішими ланцюго-утворюючими елементами сучасної мікроелектроніки. Цей метод використовує низькотемпературний лазерний скануючий мікроскоп, який досліджує фазу (тобто напрямок) та амплітуду локальних радіочастотних струмів в залежності від двовимірних координат надпровідної резонансної структури, що досліджується. Фазочутливий контраст досягається шляхом синхронізації модульованого інтенсивністю лазерного випромінювання та резонансних гармонік мікрохвильового сигналу, що проходить через зразок. У цьому випадку індуковані лазерним променем енергетичні втрати в освітлюваній області будуть сильно залежати від локальної різниці фаз між радіочастотним сигналом і просторово-часовою структурою сфокусованих осциляцій лазера. Цей підхід усуває апаратні обмеження існуючої техніки радіочастотної мікроскопії та доводить режим фазочутливої демодуляції до рівня, необхідного для вивчення фізики надпровідних метаматеріалів. Перевага представленого методу над попереднім методом радіочастотної скануючої мікроскопії демонструється на прикладі формування стоячих хвиль у спіральному надпровідному резонаторі Архімеда аж до 38-го резонансу власної моди.

Ключові слова: метаматеріал, надпровідність, спіральний резонатор, лазерний скануючий мікроскоп.